\documentclass[comunication,amsfonts,amssymb,twocolumn,showpacs]{revtex4}
\usepackage{mathrsfs}
\usepackage{graphicx,hyperref}
\usepackage{bm}
\usepackage{CJK}
\begin{document}
%\begin{CJK*}{GBK}{song}

\title{Exact control of parity-time symmetry in periodically modulated nonlinear optical couplers}
\author{Baiyuan Yang $^{1}$}
\author{Xiaobing Luo $^{1}$}
\altaffiliation{Author to whom any correspondence should be addressed: xiaobingluo2013@aliyun.com}
\author{QiangLin Hu $^{1}$}
\author{XiaoGuang Yu $^{1}$}
\affiliation{$^{1}$Department of Physics, Jinggangshan University,
Ji'an 343009, China}
\date{\today}
\begin{abstract}
We propose a mechanism  for realization of exact control of parity-time ($\mathcal{PT}$) symmetry by using
a periodically modulated nonlinear optical coupler with balanced gain and loss.
It is shown that for
certain appropriately chosen values of the modulation parameters, we can construct a family of exact analytical solutions for the two-mode equations describing the dynamics of such nonlinear couplers. These exact solutions give explicit examples that allow us to precisely manipulate the system from
nonlinearity-induced symmetry breaking to $\mathcal{PT}$ symmetry, thus providing an analytical approach
to the all-optical signal control in nonlinear $\mathcal{PT}$-symmetric structures.

\end{abstract}

\maketitle
\section{Introduction}
Enormous attentions have been attracted to Parity-time ($\mathcal{PT}$)-symmetric quantum Hamiltonian systems
 since the illustration that
their eigenvalues can be entirely real, despite the non-Hermitian Hamiltonians\cite{Bender1,Bender2,Bender3,Bender4}.
Remarkably, such
systems can exhibit a phase transition (spontaneous $\mathcal{PT}$ symmetry
breaking) from the exact to	broken-$\mathcal{PT}$ phase
whenever the gain/loss parameter exceeds
a certain threshold. In the $\mathcal{PT}$-symmetric phase, the eigenvalues of the Hamiltonian remain real and
the system exhibits periodic oscillations, while in
the $\mathcal{PT}$-broken phase, the eigenvalues become complex and the respective modes start growing exponentially.
Due to the equivalence between
 quantum mechanical Schr\"{o}dinger
equation and optical wave equation, the concept of $\mathcal{PT}$-symmetry later spreads to classical optical systems with
non-Hermitian optical potentials. Through experimental observations of $\mathcal{PT}$-symmetric optical
systems, it has been shown that such $\mathcal{PT}$-symmetric optical arrangements can exhibit unique features distinct from conservative
systems, due to non-trivial wave interference and
phase-transition effects, such as loss-induced
transparency\cite{Guo}, power unfolding and breaking of the left-right
symmetry\cite{Ruter}, unidirectional invisibility\cite{Regensburger,Feng}, optical non-reciprocity\cite{Peng}, coherent perfect laser absorbtion\cite{Sun},
and selective mode lasing in microring resonator systems\cite{Hodaei,Feng2}, and so on.

When nonlinearity is
introduced, the $\mathcal{PT}$-symmetric systems
 operating in the nonlinear regime were predicted to support additional interesting phenomena, such as solitons in
$\mathcal{PT}$ structures\cite{Wimmer}, nonlinearly induced $\mathcal{PT}$ transition\cite{Lumer}, and the all-optical
signal control\cite{Chen,Ramezani,Sukhorukov}. As was demonstrated in the study of directional
couplers with balanced gain and loss\cite{Chen,Ramezani,Sukhorukov}, nonlinear switching between $\mathcal{PT}$ symmetry
and symmetry breaking occurs above a critical
nonlinearity strength, which provides a feasible means to tailor beam shaping.
Among the extensive exploration hitherto of physics of $\mathcal{PT}$-symmetric Hamiltonians, all experiments and most of the theoretical activities have been devoted
to static (i.e., time-independent) potentials, and only very
recently growing attentions are paid to research on $\mathcal{PT}$ symmetry of time-dependent
Hamiltonians\cite{Moiseyev}-\cite{Lee}. Even so, up to now studies of time-dependent $\mathcal{PT}$-symmetric systems with nonlinearity are still rare.
Despite some isolated successes in the development
of analytical exact solution to the driven Hermitian systems with nonlinearity\cite{Deng,Xie,Hai1}, it remains elusive and very challenging to make  analytical solution of the
time-dependent $\mathcal{PT}$-symmetric systems with nonlinearity .

In this paper, we study the effect of periodic modulation on nonlinear systems with parity-time ($\mathcal{PT}$) symmetry.
By use of the appropriate harmonic parameters, we derive a family of exact solutions for the two-mode equation describing the dynamics of periodically modulated nonlinear couplers with balanced gain and loss.  These exact
solutions associated with simple modulation forms give explicit examples that enable us to analytically control the transition from
nonlinearity-induced symmetry
breaking to $\mathcal{PT}$ symmetry. In addition, these exact
solutions can give explicit illustration of $\mathcal{PT}$-symmetry analysis for the periodically modulated nonlinear system with a linear $\mathcal{PT}$-symmetric part.

\section{Exact analytical solutions and stability analysis}
\subsection{Model and $\mathcal{PT}$-symmetry properties}
We consider the simplest possible arrangement which consists
of two coupled $\mathcal{PT}$-symmetric waveguides with Kerr
nonlinearity and with the linear refractive index periodically
modulated along the propagation direction. Under coupled-mode theory,
the two modal field amplitudes are governed by the evolution equations:
\begin{eqnarray}
i\frac{d \psi_1}{dz}&=& i\gamma \psi_1+a(z)\psi_1-\chi|\psi_1|^{2} \psi_1+v \psi_2,
\nonumber\\
i\frac{d \psi_2}{dz}&=&-i\gamma \psi_2+b(z)\psi_2-\chi|\psi_2|^{2}
\psi_2+v\psi_1\nonumber\\\label{eq1}
\end{eqnarray}
with $v$ being the interchannel coupling strength,  $\gamma$ the gain or loss
strength,  $\chi$ the nonlinearity strength and the harmonic modulation being
\begin{eqnarray}\label{modulation}
a(z)&=& f \cos(\omega z+\varphi_{1}),\nonumber \\
b(z)&=& -f \cos(\omega z+\varphi_{2}).
\end{eqnarray}
Here $\psi_{1}$ and $\psi_{2}$  represent respectively field amplitudes in the first
 and the second waveguides, $\varphi_{1,2}$ denote the phases of
two harmonics, $\omega$ the modulation frequency, and $f$ the
modulation amplitude.

Throughout our paper, the dimensionless propagation distance $z$ is normalized in unit of a reference coupling length $L_c$, and propagation constant as well as all other scaled parameters $\gamma, f, \omega,v$ are in units of $L_c^{-1}$, with $L_c$ representing the shortest length for the light
switching from one waveguide to the other without nonlinearity. For instance, the reference coupling length can be selected as the one in experiment\cite{Ruter} where $L_c\approx 0.863\rm{ cm}$.  In our discussion, what is essential is the ratios between these parameters $\gamma, f, \omega,v$.

\begin{figure}[htbp]
\center
\includegraphics[width=8cm]{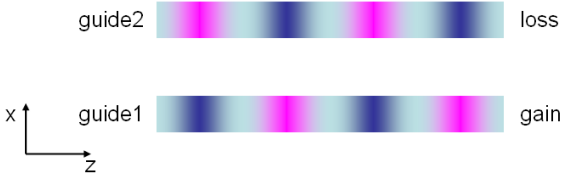}
\caption{(color online) Schematic drawing of a Kerr nonlinear $\mathcal{PT}$-symmetric coupler. The periodic
change of color along the $z$ axis denotes the periodic
modulation of the linear refractive index. One waveguide experiences
gain while
the other loss.} \label{fig1}
\end{figure}

Proceeding to the analysis of dynamics of the system (\ref{eq1}), it is instructive to identify symmetry property of the model equation.
For this purpose, we separate the Hamiltonian corresponding to the system (\ref{eq1}) into the linear part and the nonlinear part
\begin{eqnarray}\label{H}
H&=& H_L(z)+H_{NL}(\bm{\psi}(z)),\nonumber \\
H_L(z)&=&\left(
         \begin{array}{cc}
           i\gamma+a(z) & v \\
         v & -i\gamma+b(z) \\
         \end{array}
       \right),\nonumber \\
H_{NL}(\bm{\psi}(z))&=& \left(
                      \begin{array}{cc}
                        -\chi|\psi_1|^{2} & 0 \\
                        0 & -\chi|\psi_2|^{2}\\
                      \end{array}
                    \right),
\end{eqnarray}
where $\bm{\psi}(z)=(\psi_1(z), \psi_2(z))^{T}$ (hereafter the superscript $T$ stands for the matrix transpose) is the two-component vector obeying this system.

Applying the $\mathcal{PT}$ operator to both sides of equation (\ref{eq1}), where parity operator corresponds to the exchange of the two
basis states, $\mathcal{P}: \psi_1 \leftrightarrow\psi_2$, and time operator defines the reversal of propagation direction,
$\mathcal{T}: i\rightarrow -i, z\rightarrow -z$, we observe that
\begin{eqnarray}\label{trans}
\mathcal{PT}[i\partial_z \bm{\psi}(z_{0+})]&=&\mathcal{PT}\{[H_L(z_{0+})+H_{NL}(\bm{\psi}(z_{0+}))]\bm{\psi}(z_{0+})\},\nonumber \\
i\partial_z [\mathcal{PT}\bm{\psi}(z_{0+})]&=&\mathcal{PT}H_L(z_{0+})\bm{\psi}(z_{0+})\nonumber \\&&+H_{NL}(\mathcal{PT}\bm{\psi}(z_{0+}))\mathcal{PT}\bm{\psi}(z_{0+}),
\end{eqnarray}
where $z_{0\pm} :=z_0\pm z$ is defined for arbitrary constant $z_0$.
If the linear time-dependent part $H_L(z)$ is $\mathcal{PT}$ symmetric,
$\mathcal{PT}H_L(z_{0+})=H_L(z_{0+})\mathcal{PT}$,
which requires the modulation obeying the relation
\begin{eqnarray}\label{mcondi}
a(z_{0+})=b(z_{0-}),
\end{eqnarray}
then Eq.~(\ref{trans}) becomes
\begin{eqnarray}\label{trans2}
i\partial_z [\mathcal{PT}\bm{\psi}(z_{0+})]&=&H_L(z_{0+})\mathcal{PT}\bm{\psi}(z_{0+})\nonumber \\&&+H_{NL}(\mathcal{PT}\bm{\psi}(z_{0+}))\mathcal{PT}\bm{\psi}(z_{0+}).
\end{eqnarray}
By comparison of equation (\ref{trans2}) with its original equation, it is apparent that $\bm{\psi}(z_{0+})$ and $\mathcal{PT}\bm{\psi}(z_{0+})$ satisfy the same evolution equation. It also follows that the coupled-mode equation (\ref{eq1}) and the associated Hamiltonian remain invariant if
\begin{eqnarray}\label{NLcondi}
\mathcal{PT}\bm{\psi}(z_{0+})=e^{i\xi}\bm{\psi}(z_{0+}),
\end{eqnarray}
where $\xi$ is real.
The condition (\ref{NLcondi}) guarantees that
\begin{eqnarray}\label{NLcondi2}
|\psi_1(z_{0-})|=|\psi_2(z_{0+})|.
\end{eqnarray}
Actually, the requirements (\ref{NLcondi}) and (\ref{NLcondi2}) are met automatically as long as we choose
\begin{eqnarray}\label{NLcondi3}
|\psi_1(z_{0})|=|\psi_2(z_{0})|.
\end{eqnarray} Hence, this condition (\ref{NLcondi3}) is in line with the demand of $\mathcal{PT}$ symmetry for the nonlinear part.
The reasoning for derivation of (\ref{NLcondi}) and (\ref{NLcondi2}) from (\ref{NLcondi3}) is given as follows. If $\bm{\psi}(z_{0+})$ is a solution of Eq.~(\ref{eq1}), then $e^{i\xi}\bm{\psi}(z_{0+})$ and $\mathcal{PT}\bm{\psi}(z_{0+})$ are also solution of Eq.~(\ref{eq1}). Suppose that $\mathcal{PT}\bm{\psi}(z_{0})=e^{i\xi}\bm{\psi}(z_{0})$, which can be achieved by setting $|\psi_1(z_{0})|=|\psi_2(z_{0})|$ with an appropriately chosen value of phase $\xi$. It follows that (\ref{NLcondi}) and (\ref{NLcondi2}) hold, since $e^{i\xi}\bm{\psi}(z_{0+})$ and $\mathcal{PT}\bm{\psi}(z_{0+})$ satisfy the same evolution equation.
When the conditions (\ref{mcondi}) and (\ref{NLcondi3}) are satisfied simultaneously, the Hamiltonian (\ref{H}) for such a periodically modulated nonlinear coupler with balanced gain and loss is invariant under the combined action of parity operator $\mathcal{P}$  and time operator $\mathcal{T}$.
Therefore, for the system (\ref{eq1}) to be $\mathcal{PT}$ symmetric, the linear time-dependent part $H_L(z)$ must be $\mathcal{PT}$ symmetric, which demands that the condition ({\ref{mcondi}}) be satisfied and the gain and loss coefficient be taken equal, as well as the nonlinear part fulfill the condition (\ref{NLcondi3}) at the same $z_0$. However, for our model, the conditions ({\ref{mcondi}}) and (\ref{NLcondi3}) are not sufficient for the $\mathcal{PT}$ symmetry to be unbroken.
The $\mathcal{PT}$ symmetry is said to be unbroken if and only if there exist $\mathcal{PT}$-invariant Floquet eigenstates with real quasienergies.

In considering the propagation of a light wave with
electric field $E(x, z)$ along the $z$ axis in a nonlinear
setup, the complex refractive index (playing the role of
an optical potential) of the Kerr-type medium is written as
$n(x,z)=n_R(x,z)+in_I(x)-n_2|E(x,z)|^2$ , where $n_2$ is the nonlinear refractive index.  In our case, we only consider
an Hermitian driving, that is, $n_I$ does not depend on time ($z$). For the system to be symmetric, $n(x,z)$ must be symmetric with respect to combined
action of the parity $\mathcal{P}$ and time-reversal $\mathcal{T}$ operators, i.e., $n^*(-x,z_0-z)=n(x,z_0+z)$ for arbitrary constant $z_0$, which demands that $n_R(-x,z_{0-})=n_R(x,z_{0+}),n_I(-x)=-n_I(x), |E(-x,z_{0})|=|E(x,z_{0})|$.
As we can see, the imaginary part still must be an odd function of position, as is common for the linearly equivalent $\mathcal{PT}$-symmetric system.
For the non-driven limit
(periodic driving is turned off) of nonlinear
model that we are discussing, $\mathcal{PT}$ symmetry requires that the intensity distribution possesses spacetime-reflection symmetry, apart from the linear $\mathcal{PT}$-symmetry prerequisites (the symmetric real part and the antisymmetric imaginary
part of the linear index of refraction), since the nonlinearity has to be regarded as a contribution
to the optical potential\cite{Li,Dast}.

\subsection{Exact Floquet solutions}
The nonlinear $\mathcal{PT}$-symmetric coupler was introduced in Refs.\cite{Chen,Ramezani,Sukhorukov}, where the interplay between $\mathcal{PT}$ symmetry with nonlinear effects was discussed through the symmetry analysis as well as construction of
exact analytical solution describing the nonlinear dynamics. When the periodic modulation is added,
the situation is complicated and finding exact analytical solutions for such a problem becomes extremely hard. However, here we will show that a family of exact
analytical solutions exists if given with certain suitably chosen
values of the modulation parameters.
The exact analytical solution of Eq.~(\ref{eq1}) can be explicitly written as
\begin{eqnarray}\label{S}
\psi_{1}(z)&=&e^{-i[\frac{\omega}{2}-\chi(|C_{1}|^{2}+|C_{2}|^{2})] z}(C_{1}e^{i\vartheta}-C_{2}e^{-i\vartheta+i\omega z}),
\nonumber \\
\psi_{2}(z)&=& e^{-i[\frac{\omega}{2}-\chi(|C_{1}|^{2}+|C_{2}|^{2})] z}(C_{1}+C_{2}e^{i\omega z}),
\end{eqnarray}
in the case of $\gamma<v$, that is, the values of gain and loss
coefficient below the linear $\mathcal{PT}$-symmetry-breaking threshold,
by adjusting the modulation parameters to satisfy the following conditions
\begin{eqnarray}\label{PR}
 f &=& -2\chi|C_{1}||C_{2}|, \\
  \omega &=&2v \cos\vartheta, \\
  \varphi_{1} &=& -2\vartheta+(\beta_{2}-\beta_{1}), \\
\varphi_{2}  &=& \beta_{2}-\beta_{1},\\
\sin\vartheta &=&\gamma/v,\\
\cos\vartheta&=&\sqrt{1-(\gamma/v)^2},\label{PR2}
\end{eqnarray}
where $C_{1}$ and $C_{2}$ are constants dependent on initial conditions $\psi_{1,2}(0)$,
\begin{eqnarray}
C_{1}&=&|C_{1}|e^{i\beta_{1}}=\frac{\psi_{1}(0)+\psi_{2}(0)e^{-i\vartheta}}{2\cos\vartheta},\label{C1}\\
C_{2}&=&|C_{2}|e^{i\beta_{2}}=\frac{-\psi_{1}(0)+\psi_{2}(0)e^{i\vartheta}}{2\cos\vartheta}.\label{C2}
\end{eqnarray}
According to the Floquet theorem, this solution is a special kind of
nonlinear Floquet state in the form of $(\psi_1(z), \psi_2(z))^{T}=\bm{\Phi}(z)e^{-i\varepsilon z}=(\Phi_1(z), \Phi_2(z))^{T}e^{-i\varepsilon z}$,
where the Floquet eigenmodes $\bm{\Phi}(z)$ are periodic with the modulation period $\Lambda=2\pi/\omega$ and
the corresponding quasienergy $\varepsilon=\omega/2-\chi(|C_{1}|^{2}+|C_{2}|^{2})$. From Eq.~(\ref{S}) it can be observed that the exact solution is periodic because the corresponding quasienergy is entirely real despite a non-Hermitian
Hamiltonian. The exact solutions of Eq.~(\ref{S})
can be
directly proved by substituting them into Eq.~(\ref{eq1}) with the help of Eqs.~(\ref{PR})-(\ref{C2}).

It can also be easily verified that this exact solution obeys the following relation\cite{Bronski,Hai}
\begin{eqnarray}\label{BL}
a(z)-\chi|\psi_{1}|^{2}=b(z)-\chi|\psi_{2}|^{2}=-\chi(|C_{1}|^{2}+|C_{2}|^{2}),
\end{eqnarray}
which indicates that the optical nonlinear effect is balanced by the external modulation. Under the balanced condition (\ref{BL}), equation (\ref{eq1}) is reduced to the linear Schr\"{o}dinger equation
\begin{eqnarray}
i\frac{d \psi_1}{dz}&=& i\gamma \psi_1-\chi(|C_{1}|^{2}+|C_{2}|^{2})\psi_1+v \psi_2,
\nonumber\\
i\frac{d \psi_2}{dz}&=&-i\gamma \psi_2-\chi(|C_{1}|^{2}+|C_{2}|^{2})
\psi_2+v\psi_1.\nonumber\\\label{eq2}
\end{eqnarray}
The exact Floquet solution (\ref{S}) can be obtained with assumption that a solution to the system (\ref{eq1}) containing undetermined modulation  parameters obeys
the balance condition (\ref{BL}) and linear equation
(\ref{eq2}) simultaneously.
It should be noted that our exact solutions can be constructed only in the linear unbroken $\mathcal{PT}$-symmetric phase ($\gamma<v$).
At the exceptional point ($\gamma=v$) or in the broken $\mathcal{PT}$-symmetric phase ($\gamma>v$), the solution of linear equation
(\ref{eq2}) is unbounded such that the mode intensity can not balance the harmonic (periodic) modulation and the preestablished condition (\ref{BL}) not be
satisfied, indicating that the method we use in constructing exact analytical solutions is not applicable in these regimes. At present, it is still very challenging for us to find stable nonlinear Floquet eigenmodes with real quasienergies in the domains of broken $\mathcal{PT}$ symmetry ($\gamma>v$). This open problem deserves further study.

As is demonstrated in
 Refs.~\cite{Chen,Ramezani,Sukhorukov}, for the unmodulated nonlinear coupler with balanced gain and loss, if the nonlinearity strength $\chi$ is small, the beam evolution will show periodic oscillation ($\mathcal{PT}$-symmetric phase), whereas above a critical nonlinearity strength, nonlinearity will lead to symmetry-breaking and exponential growth of beam power. When the harmonic modulation parameters are appropriately chosen to satisfy conditions (\ref{PR})-(\ref{C2}), it is possible to move the system to periodic state (\ref{S}) for arbitrary value of nonlinearity strength.

 Let us take a closer look at the properties of Floquet states. The Floquet theorem tells us that if the Hamiltonian (\ref{H}) depends periodically on time ($z$), there exists a set of Floquet solutions to the time-dependent  Schr\"{o}dinger equation (\ref{eq1}). However, if the quasi-energy is complex, the nonlinear part of Hamiltonian $H_{NL}$ will not be periodic with the mode intensities either exponentially increasing or decaying, showing no existence of such a Floquet solution with complex quasi-energy. Substituting the Floquet solution $\bm{\psi}(z)=\bm{\Phi}(z)e^{-i\varepsilon z}$ into Eq.~(\ref{eq1}), we obtain the following eigenvalue equation\cite{Sambe}
 \begin{eqnarray}\label{eigenvalue}
[H(z)-i\partial_z]\bm{\Phi}(z)=\varepsilon\bm{\Phi}(z),\\
H(z)=H_L(z)+H_{NL}(\bm{\Phi}(z)),  \nonumber
\end{eqnarray}
where $\rm{Im}(\varepsilon)=0$
is required for the existence of nonlinear Floquet eigenmodes, and $H_{NL}(\bm{\psi}(z))=H_{NL}(\bm{\Phi}(z))$ is utilized, thanks to the phase invariance of nonlinearity. The operator $\mathcal{H}:=H(z)-i\partial_z$ is the so-called Floquet Hamiltonian in the
extended Hilbert space \cite{Sambe} and $\bm{\Phi}(z)$ is called Floquet eigenstate (eigenmode). For the eigenvalue problem, the $z$ (time) periodicity of
the Floquet Hamiltonian, $\mathcal{H}(z+2\pi/\omega)=\mathcal{H}(z)$, is required for ensuring that $\bm{\Phi}(z+2\pi/\omega)=\bm{\Phi}(z)$.
Acting with $\mathcal{PT}$ on Eq.~(\ref{eigenvalue}), we have
 \begin{eqnarray}\label{PT2}
\mathcal{PT}\{[H(z_{0+})-i\partial_z]\bm{\Phi}(z_{0+})\}=\mathcal{PT}[{\varepsilon\bm{\Phi}(z_{0+})}],\nonumber\\
\{H_L(z_{0+})+H_{NL}(\mathcal{PT}\bm{\Phi}(z_{0+}))-i\partial_z\}\mathcal{PT}\bm{\Phi}(z_{0+})\nonumber\\=\varepsilon\mathcal{PT}\bm{\Phi}(z_{0+}),\nonumber\\
\end{eqnarray}
where ({\ref{mcondi}}) and $[H_L(z_{0+}),\mathcal{PT}]=0$ are used.  For non-degenerate real eigenvalues,
(\ref{PT2}) requires that the Floquet eigenstate is $\mathcal{PT}$-symmetric, $\mathcal{PT}\bm{\Phi}(z_{0+})=e^{i\delta}\bm{\Phi}(z_{0+})$, where $\delta$ is real, because $\bm{\Phi}(z_{0+})$ and $\mathcal{PT}\bm{\Phi}(z_{0+})$
obey the same eigenvalue equation. In such case, the Floquet Hamiltonian operator $\mathcal{H}$ and the $\mathcal{PT}$ operator share the same eigenmode. As a matter of fact, occurrence of $\mathcal{PT}$-symmetric eigenstate with real eigenvalue (unbroken $\mathcal{PT}$ symmetry) is the most striking feature of $\mathcal{PT}$ symmetry. When we vary the system parameters, the eigenvalue equation (\ref{eigenvalue}) itself may possess physically distinct eigenmodes $\bm{\Phi}(z_{0+})$ and $\mathcal{PT}\bm{\Phi}(z_{0+})$ corresponding to pairwise complex quasienergies $\varepsilon$ and $\varepsilon^*$, which can be viewed as a manifestation of $\mathcal{PT}$ symmetry breaking\cite{Lumer}.
Though the system (\ref{eq1}) has unbounded solution due to $\mathcal{PT}$ symmetry breaking, we emphasize again that such broken $\mathcal{PT}$-symmetric Floquet eigenstates with complex quasienergies do not occur in the considered system.  This is in full analogy to the fact that there exist only stationary nonlinear modes with real eigenvalues for the corresponding unmodulated system\cite{Konotop,Konotop2}.

We can readily check that our exact Floquet eigenstates in
Eq.~(\ref{S}) are $\mathcal{PT}$-invariant (up to trivial phase shift), that is, for $z_0=[(2k+1)\pi-\varphi_2-\varphi_1]/(2\omega)=[(2k+1)\pi+2\vartheta-2(\beta_2-\beta_1)]/(4v\cos\vartheta)$ with $k$ being integer,
 \begin{eqnarray}\label{PTst}
\mathcal{PT}\bm{\Phi}(z_{0+}) &=& \mathcal{PT}\left(
                                             \begin{array}{c}
                                              C_{1}e^{i\vartheta}-C_{2}e^{-i\vartheta+i\omega (z_0+z)} \\
                                             C_{1}+C_{2}e^{i\omega (z_0+z)} \\
                                             \end{array}
                                           \right)\nonumber\\
                                            &=&e^{-i(2\beta_1+\vartheta)}\bm{\Phi}(z_{0+}).
\end{eqnarray}
It is also observed that for a given exact solution (\ref{S}), the conditions (\ref{mcondi}) and (\ref{NLcondi3}) for the Hamiltonian to
be $\mathcal{PT}$ symmetric are naturally satisfied.

Eqs.~(\ref{S})-(\ref{C2}) comprise the central result of this paper. Every harmonic modulation (\ref{modulation}) satisfying Eqs.~(\ref{PR})-(\ref{C2}) will generate an explicit example that
allows for exactly controlling $\mathcal{PT}$ symmetry transition.  We will now demonstrate this by taking several specific cases as
examples.

Case 1: We first consider the exact control of $\mathcal{PT}$ symmetry where the system is excited at either channel 1 or channel 2.
Applying the initial condition $\psi_{1}(0)=1, \psi_{2}(0)=0$ to Eqs.~(\ref{C1}) and (\ref{C2}), and utilizing (\ref{PR})-(\ref{PR2}), we give the explicit form of harmonic modulations as follows
\begin{eqnarray}\label{M1}
    a(z)&=&\frac{\chi\cos(2v \cos\vartheta z-2\vartheta)}{2\cos^{2}\vartheta},\nonumber \\
    b(z)&=&-\frac{\chi\cos(2v \cos\vartheta z)}{2\cos^{2}\vartheta}.
\end{eqnarray}
 Given initial input beam $\psi_{1}(0)=1, \psi_{2}(0)=0$, we thus determine the
shape of harmonic modulation, given by Eq.~(\ref{M1}), for which the system (\ref{eq1}) can be exactly solved.
Under such circumstance, the light intensities at the first and the second waveguides exactly read as
\begin{eqnarray}
|\psi_{1}(z)|^{2}&=&\frac{1}{2\cos^{2}\vartheta}+\frac{\cos(2v \cos\vartheta z-2\vartheta)}{2\cos^{2}\vartheta},\nonumber \\ |\psi_{2}(z)|^{2}&=&\frac{1}{2\cos^{2}\vartheta}-\frac{\cos(2v  \cos\vartheta z)}{2\cos^{2}\vartheta}.\label{iniP1}
\end{eqnarray}
For the exact solution (\ref{iniP1}) associated with the initial condition $\psi_{1}(0)=1, \psi_{2}(0)=0$, it is easily verified that
at $z_0=[(2k+1)\pi+2\vartheta]/(4v\cos\vartheta)$, the relations (\ref{mcondi}) and (\ref{NLcondi3}) hold, i.e.,
 \begin{eqnarray}\label{PTv}
 a(z_0+z)&=& b(z_0-z)=\frac{(-1)^k\chi\sin(\vartheta-2v\cos\vartheta z)}{2\cos^{2}\vartheta},\nonumber\\
 |\psi_{1}(z_0)|^{2}&=& |\psi_{2}(z_0)|^{2}=\frac{1}{2\cos^{2}\vartheta}+\frac{(-1)^k\sin\vartheta}{2\cos^{2}\vartheta}.
\end{eqnarray}
Since the time evolution of the total light intensity, $I=|\psi_{1}|^2+|\psi_{2}|^2$, is determined by
\begin{equation}\label{Norm}
  \frac{dI}{dz}=2\gamma ( |\psi_{1}(z)|^{2}-|\psi_{2}(z)|^{2}),
\end{equation}
we conclude that $dI/dz|_{z=z_0}=0$, and thus the total intensity reaches the maxima or minima at $z_0=[(2k+1)\pi+2\vartheta]/(4v\cos\vartheta)$ with $k$ being integer.

For another set of initial values $\psi_{1}(0)=0, \psi_{2}(0)=1$, Eqs.~(\ref{PR})-(\ref{C2}) ) give the explicit form of harmonic modulations
\begin{eqnarray}\label{M2}
    a(z)&=&-\frac{\chi\cos(2v \cos\vartheta z)}{2\cos^{2}\vartheta},\nonumber \\
    b(z)&=&\frac{\chi\cos(2v \cos\vartheta z+2\vartheta)}{2\cos^{2}\vartheta}.
\end{eqnarray}
The shape of modulation (\ref{M2}), associated with the initial input $\psi_{1}(0)=0, \psi_{2}(0)=1$, produces the exact analytical distributions of light as
\begin{eqnarray}
    |\psi_{1}(z)|^{2}
                  &=&\frac{1}{2\cos^{2}\vartheta}-\frac{\cos(2v \cos\vartheta z)}{2\cos^{2}\vartheta},\nonumber \\
     |\psi_{2}(z)|^{2}
                  &=&\frac{1}{2\cos^{2}\vartheta}+\frac{\cos(2v  \cos\vartheta z+2\vartheta)}{2\cos^{2}\vartheta}.\label{iniP2}
\end{eqnarray}

Similarly, for the exact solution (\ref{iniP2}), we can prove that at  $z_0=[(2k+1)\pi-2\vartheta]/(4v\cos\vartheta)$ with $k$ being integer,
the $\mathcal{PT}$ symmetry requirement (\ref{mcondi}) and (\ref{NLcondi3}) are also met, i.e.,
\begin{eqnarray}\label{PTv}
 a(z_0+z)&=& b(z_0-z)=\frac{(-1)^{k+1}\chi\sin(\vartheta-2v\cos\vartheta z)}{2\cos^{2}\vartheta},\nonumber\\
 |\psi_{1}(z_0)|^{2}&=& |\psi_{2}(z_0)|^{2}=\frac{1}{2\cos^{2}\vartheta}+\frac{(-1)^{k+1}\sin\vartheta}{2\cos^{2}\vartheta}.
\end{eqnarray}

\begin{figure}[htbp]
\center
\includegraphics[width=8cm]{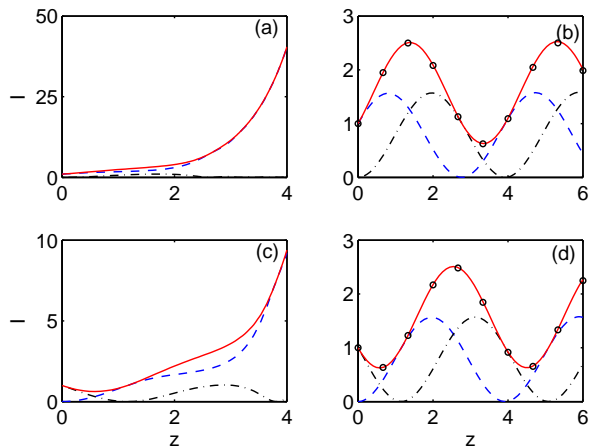}
\caption{(color online) Beam dynamics of the coupled-mode equation (\ref{eq1}) for two different initial conditions:
(a)-(b) $\psi_{1}(0)=1, \psi_{2}(0)=0$ and (c)-(d) $\psi_{1}(0)=0, \psi_{2}(0)=1$. (a) and (c) are for the no-modulation case; (b) and (d) are for the case with modulation. Shown
here are the numerical results of intensity $I_1 =|\psi_1|^2 $ in the first waveguide
(blue dashed line), $I_2 =|\psi_1|^2 $ in the second waveguide (black
dash-dotted line), and the total intensity $I = I_1 +I_2$ (red solid
line). The open circles denote the exact analytical results of the total intensity. The system parameters are chosen as $v=1, \gamma=0.6, \chi=1.2$, and the modulation parameters for (b) and (d) are given by Eq.~(\ref{M1}) and Eq.~(\ref{M2}), respectively.} \label{fig2}
\end{figure}

We present in Fig.~\ref{fig2} two examples of exact control of transition between symmetry breaking and $\mathcal{PT}$ symmetry
with controlled harmonic modulations. In our simulation, the numerical solutions of the coupled-mode equation (\ref{eq1}) are obtained with the use of Runge-Kutta method. Figs.~\ref{fig2}(a)-(b) show the intensity evolution for the initial state $\psi_{1}(0)=1, \psi_{2}(0)=0$.
For the no-modulation case [Fig.~\ref{fig2}(a)], where the
system parameters are given by $ v=1, \gamma=0.6, \chi=1.2, f=0$, we observe that the total intensity
exhibits exponential growth with light concentrating in the waveguide with gain, corresponding to a nonlinearly induced symmetry breaking.
Fig.~\ref{fig2}(b) indicates completely different dynamics, where the light periodically
oscillates when the Eq.~(\ref{M1})-obeying periodic modulation is applied, which indicates a $\mathcal{PT}$ symmetric state.
Likewise, Figs.~\ref{fig2}(c)-(d) represent the exact control from
nonlinearity-induced symmetry breaking [Fig.~\ref{fig2}(c)] to $\mathcal{PT}$ symmetry [Fig.~\ref{fig2}(d)] for another initial state $\psi_{1}(0)=0, \psi_{2}(0)=1$ by performing the harmonic modulation in the form of Eq.~(\ref{M2}). As is apparent in  Fig.~\ref{fig2}, the modulations employed for exact control of $\mathcal{PT}$ symmetry are distinct and light evolution is obviously non-reciprocal by exchanging the input channel from
1 to 2. We have compared the analytical (circles) with the numerical (red solid
lines) results in Figs.~\ref{fig2} (b) and (d). The agreement is almost perfect.

Case 2: We now turn to the case of  $\psi_{1}(0)=\psi_{2}(0)=1/\sqrt{2}$ and consider $-\infty < z<+\infty$, which was discussed in Ref.~\cite{Sukhorukov}.  Take our harmonic modulation as
 \begin{eqnarray}\label{M3}
    a(z)&=&-\frac{\chi\sin\vartheta\cos(2v \cos\vartheta z-\vartheta+\frac{\pi}{2})}{2\cos^{2}\vartheta},\nonumber \\
    b(z)&=&\frac{\chi\sin\vartheta\cos(2v \cos\vartheta z+\vartheta+\frac{\pi}{2})}{2\cos^{2}\vartheta}.
\end{eqnarray}
For the initial condition $\psi_{1}(0)=\psi_{2}(0)=1/\sqrt{2}$, the analytical distribution of light intensity corresponding to the modulation (\ref{M3})
is
 \begin{eqnarray}
    |\psi_{1}(z)|^{2}
                  &=&\frac{1}{2\cos^{2}\vartheta}+\frac{\sin\vartheta\sin(2v \cos\vartheta z-\vartheta)}{2\cos^{2}\vartheta},\nonumber \\
     |\psi_{2}(z)|^{2}
                  &=&\frac{1}{2\cos^{2}\vartheta}-\frac{\sin\vartheta\sin(2v  \cos\vartheta z+\vartheta)}{2\cos^{2}\vartheta}.\label{iniP3}
\end{eqnarray}
\begin{figure}[htbp]
\center
\includegraphics[width=8cm]{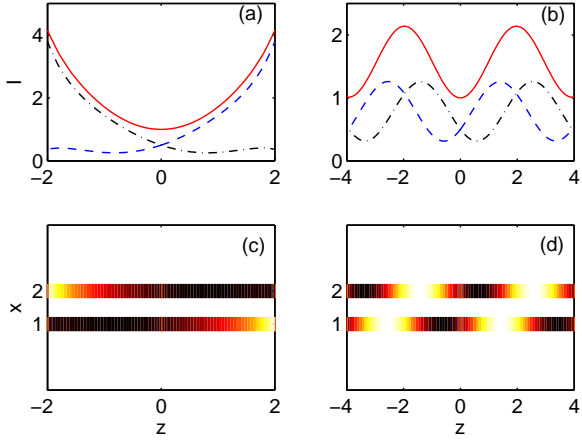}
\caption{(color online) Numerical solution of the coupled
equations (\ref{eq1}) describing the nonlinear $\mathcal{PT}$-symmetric system with the initial condition $\psi_{1}(0)=\psi_{2}(0)=1/\sqrt{2}$.
(a) and (c) are for the case with no modulation; (b) and (d) are for the case with modulation (\ref{M3}).
(a)-(b) Dependencies of intensity on
propagation distance $z$ in the first (blue dashed line) and second (black dash-dotted line)
waveguides; the solid line shows the total intensity. (c) and (d) give visualization of the intensity evolution along the
propagation direction. In all plots, $ v=1,\gamma=0.6$, the nonlinearity strength $\chi=1.2$, and the modulation parameters are given by Eq.~(\ref{M3}).}\label{fig3}
\end{figure}

Reported in Fig.~\ref{fig3} is the beam dynamics associated with the initial condition $\psi_{1}(0)=\psi_{2}(0)=1/\sqrt{2}$.
As above, system parameters are fixed as $ v=1, \gamma=0.6, \chi=1.2$, where the unmodulated system lies in the region of symmetry breaking with nonlinear switching. In fact, this result is confirmed
by numerical simulations in Figs.~\ref{fig3} (a) and (c), where the
output beam leaves the sample from the waveguide
with gain and the total intensity increases without bound.
In contrast, when the harmonic modulation (\ref{M3}) is turned on, the periodic intensity oscillations are observed in Figs.~\ref{fig3} (b) and (d).
Another nontrivial observation is that the dynamics starting from
$\psi_{1}(0)=\psi_{2}(0)=1/\sqrt{2}$ in positive ($+z$) and negative ($-z$) directions is exactly
equivalent, upon exchanging the two channels labeled by 1
and 2.
\subsection{Stability analysis}
In this subsection, the stability of these exact solutions will be discussed.
For this purpose we first define
the renormalized wave function with components
 \begin{equation}\label{Renor}
    \phi_{1}(z)=\frac{\psi_{1}(z)}{\sqrt{I}}, \phi_{2}(z)=\frac{\psi_{2}(z)}{\sqrt{I}}.
\end{equation}

The dynamics of system (\ref{eq1}) thus can be expressed in terms of the renormalized wave function
\begin{eqnarray}
i\frac{d \phi_1}{dz}&=& i\gamma (1-\kappa)\phi_1+a(z)\phi_1-\chi I|\phi_1|^{2} \phi_1+v \phi_2,
\nonumber\\
i\frac{d \phi_2}{dz}&=&-i\gamma(1+\kappa) \phi_2+b(z)\phi_2-\chi I|\phi_2|^{2}
\phi_2+v\phi_1,\nonumber\\\label{N1}
\end{eqnarray}
with $\kappa=|\phi_1|^{2}-|\phi_2|^{2}$. This dynamics by definition conserves
the normalization $|\phi_1|^{2}+|\phi_2|^{2}=1$.
It will be convenient
to define three real
components of a Bloch vector with respect to the renormalized wave function
\begin{eqnarray}\label{Bloch}
  S_{x} &=& \frac{\phi_{1}^{\ast}\phi_{2}+\phi_{1}\phi_{2}^{\ast}}{2}, \nonumber \\
  S_{y} &=& \frac{\phi_{1}^{\ast}\phi_{2}-\phi_{1}\phi_{2}^{\ast}}{2 i}, \nonumber \\
  S_{z} &=& \frac{|\phi_{1}|^{2}-|\phi_{2}|^{2}}{2}.
\end{eqnarray}
In this representation, we obtain the generalized Bloch
equations of motion from (\ref{N1}) as
\begin{eqnarray}\label{EM}
  \dot{S}_{x} &=& [b(z)-a(z)]S_{y} -4\gamma S_{x}S_{z}+2\chi IS_{y}S_{z}, \nonumber\\
  \dot{S}_{y} &=& [a(z)-b(z)]S_{x} -2v S_{z}-2\chi IS_{x}S_{z}-4\gamma S_{y}S_{z}, \nonumber \\
  \dot{S}_{z} &=& \gamma +2v S_{y}-4\gamma S_{z}^{2}.
  \end{eqnarray}
These Bloch equations conserve $S_{x}^2+S_{y}^2+S_{z}^2=1/4$, which means that the evolutions of variables $S_{x},S_y$ and $S_z$
are confined to the Bloch sphere. Besides, the total intensity $I$ evolves as
\begin{equation}\label{Norm}
  \dot{I}=4\gamma I{S}_{z}.
\end{equation}
If we consider a small perturbation $(\delta S_{x},\delta S_{y},\delta S_{z},\delta I)$ from the exact solution, a linearized equation is found for the perturbations
\begin{eqnarray}\label{de}
\dot{\delta S_{x}} &=&-4\gamma S_{z}\delta S_{x}+[-a(z)+b(z)+2\chi IS_{z}]\delta S_{y}\nonumber\\
                         &&-(4\gamma S_{x}-2\chi IS_{y})\delta S_{z}+2\chi S_{y}S_{z} \delta I,\nonumber\\
\dot{\delta S_{y}}&=& [a(z)-b(z)-2\chi IS_{z}]\delta S_{x}-4\gamma S_{z}\delta S_{y}\nonumber\\
                        &&-(2\nu+2\chi IS_{x}+4\gamma S_{y})\delta S_{z}-2\chi S_{x}S_{z} \delta I,\nonumber \\
\dot{\delta S_{z}} &=& 2\nu \delta S_{y}-8\gamma S_{z}\delta S_{z},\nonumber \\
\dot{\delta I}&=&4\gamma I\delta S_{z}+4\gamma S_{z}\delta I.
\end{eqnarray}

Eq.~(\ref{de}) constitutes a set of linear differential equations
with periodic coefficients, which admit solutions of the form  $V(z)=e^{\lambda z}u(z)$, where $\lambda$ is often
called quasienergy and $u(z)$ is the Floquet function periodic with the period
of the coefficients in equation (\ref{de}). These exact solutions are linearly stable if and only if
there is no quasienergy with a positive real part.  To find the quasienergies we adopt the numerical
technique similar to that given in Ref.~\cite{Zheng}.  We first numerically integrate
the linear differential equation $\ref{de}$ from $z = 0$ to $\Lambda=2\pi/\omega$ with the initial conditions $[1,0,0,0]^T,[0,1,0,0]^T,
[0,0,1,0]^T,[0,0,0,1]^T$ to obtain the results $[U_{j,1},U_{j,2},U_{j,3},U_{j,4}]^{T},j=1,2,3,4$, respectively. The resulting matrix $U$ is the time evolution operator of Eq.~(\ref{de}) over one modulation period. Given that the Floquet functions $u(z)$ are eigenstates of $U$ with eigenvalues
$\exp(\lambda \Lambda)$, the quasienergies of the linear equation (\ref{de}) are numerically computed
by direct diagonalization of $U$.

\begin{figure}[htbp]
\center
\includegraphics[width=8cm]{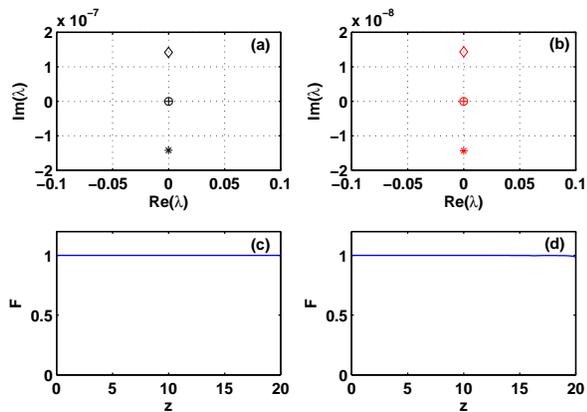}
\caption{(color online) (a)-(b) Quasienergies $\lambda$ of the linearized equation (\ref{de}) corresponding to two different exact solutions. (a) and (b) are for the quasienergies $\lambda$ associated with
two exact solutions shown in Figs.~\ref{fig2} (b) and (d), respectively.
(c)-(d) the evolutions of the fidelity between the exact solution  and
the numerical solution.  Here, (c) represents the fidelity between the exact solution shown in Fig.~\ref{fig2} (b) and its corresponding numerical solution with initial random noise. (d) denotes the fidelity between the exact solution shown in Fig.~\ref{fig2} (d) and the numerical one with initial random noise.}\label{fig4}
\end{figure}

In Figs.~(\ref{fig4}) (a)-(b), we exhibits the numerically obtained quasienergies $\lambda$ of the linearized equations related to the two exact solutions shown in Figs.~\ref{fig2} (b) and (d). These two exact
solutions, one of which corresponds to the initial condition $\psi_{1}(0)=1, \psi_{2}(0)=0$ and the other to $\psi_{1}(0)=0, \psi_{2}(0)=1$, are dynamically stable, because their linearized equations have no quasienergy with a positive real part. In fact, all the quasiengies displayed in Figs.~\ref{fig4} (a) and (b) have $\rm{Re}(\lambda)=0$, in which case the perturbations are purely oscillating. To double check the stability results, we calculate the difference between the exact solution without perturbations and
the numerical one with initial random noise, which is characterized by a dynamical fidelity
\begin{equation}\label{Fid}
  F(z)=\frac{1}{\mathcal{N}}|\psi_{1,\rm{num}}^{*}\psi_{1,\rm{ex}}+\psi_{2,\rm{num}}^{*}\psi_{2,\rm{ex}}|^2
\end{equation}
with normalization function defined as
\begin{equation}\label{Fid}
 \mathcal{N}=(|\psi_{1,\rm{num}}|^2+|\psi_{2,\rm{num}}|^2)(|\psi_{1,\rm{ex}}|^2+|\psi_{2,\rm{ex}}|^2).
\end{equation}
Here,
the functions $\psi_{1,\rm{num}}$ and $\psi_{2,\rm{num}}$ denote the two modal field
amplitudes of numerical solutions of system (\ref{eq1}) with initial perturbation added to the exact solution,
$\psi_{1,\rm{ex}}$ and $\psi_{2,\rm{ex}}$ the two components of exact solution, and
superscript $*$ the complex conjugate. The normalization function
ensures that $0\leq F(z)\leq 1$.
The fidelity $F(z)$ is a measure of the similarity between the exact solution and the numerical solution:
when $F =1$, the two solutions are the same; when $F=0$, they have no similarity.
We present the evolutions of the fidelity between the exact solution  and
the numerical solution for two cases, as shown in Fig.~\ref{fig4}(c)-(d) . In Fig.~\ref{fig4}(c), the exact solution is same as in Fig.~\ref{fig2} (b), where the harmonic modulation is given by (\ref{M1}) and the initial state is $\psi_{1}(0)=1, \psi_{2}(0)=0$;  the numerical solution is obtained through integration of Eq.~(\ref{eq1}) with the same modulations and
parameters as those in Fig.~\ref{fig2} (b) except for random perturbations added to the initial state $\psi_{1}(0)=1, \psi_{2}(0)=0$.
 In Fig.~\ref{fig4}(d), the exact solution corresponds to the one shown in Fig.~\ref{fig2} (d); the numerical solution is obtained via integration of Eq.~(\ref{eq1}) in the presence of random perturbations to the initial state $\psi_{1}(0)=0, \psi_{2}(0)=1$  for the same modulations and
parameters as those in Fig.~\ref{fig2} (d).
As clearly seen in Fig.~\ref{fig4}(c)-(d), for both cases, the fidelities $F(z)$ perfectly
remain unity, which confirms that the two exact solutions shown in Figs.~\ref{fig2} (b) and (d) are stable against the numerical perturbation.  Moreover, our numerical results (not displayed here) show that all the exact solutions associated with the initial conditions $\psi_{1}(0)=1, \psi_{2}(0)=0$ and $\psi_{1}(0)=0, \psi_{2}(0)=1$ are stable for the system parameters $\gamma/v$ in the range of $0\leq\gamma/v<1$.

\section{Conclusion}
In conclusion, we have studied analytically the effect of periodic modulation on
$\mathcal{PT}$-symmetry of nonlinear directional waveguide couplers with balanced
gain and loss. A class of explicit analytical solution for the coupled-mode equation describing the dynamics of such nonlinear couplers has been constructed, which gives unlimited number of examples for analytical demonstration of the exact control of $\mathcal{PT}$-symmetry transition.  With suitably chosen parameters of harmonic modulation, we can exactly control the system from nonlinearity-induced symmetry breaking to $\mathcal{PT}$-symmetry, which provides an analytical approach to tailor beam shaping and switching in nonlinear $\mathcal{PT}$-symmetric
lattices.  The
technique used here to get the exactly solvable control fields can be applied in other physical contexts such as Bose-Einstein condensates in a $\mathcal{PT}$-symmetric double well.

The work was supported by the NSF of China under Grants 11465009,
11165009, the Doctoral Scientific Research Foundation of jinggangshan university (JZB15002), and the Program for New Century Excellent Talents in
University of Ministry of Education of China (NCET-13-0836).

%\end{CJK*}

\end{document}